\documentclass[a4paper,11pt]{article}
\usepackage[a4paper]{geometry}
\usepackage{lineno,hyperref} 

\advance\textheight by 2cm
\advance\topmargin -1cm
\advance\textwidth by 1.5cm
\advance\oddsidemargin by -0.75cm
\advance\evensidemargin by -0.75cm

\newtheorem{definition}{Definition}
\newtheorem{proposition}{Proposition}

\newcounter{noqed}
\newcommand{\qed}{ \ifmmode\mbox{ }\fi\rule[-.05em]{.3em}{.7em}\setcounter{noqed}{0}}
\newenvironment{proof}[1][{}]{\noindent{\bf Proof#1.}}{\qed\par\medskip}

\newcommand{\HD}{H\"{o}rmann--Derflinger\xspace}
\newcommand{\sqrtd}{\sqrt[d\mskip-4.5mu]}   

\usepackage{booktabs}
\usepackage{pdfsync}
\usepackage{verbatim}
\usepackage{graphicx}
\usepackage{xspace}
\usepackage{mathrsfs}
\usepackage{amsmath}
\usepackage{amssymb}
\usepackage{multirow}
\usepackage{rotating}
\usepackage{booktabs}
\usepackage{bm}
\usepackage{url}

\def\..{\,\mathpunct{\ldotp\ldotp}} 

\newcommand{\Z}{\mathbf Z}

\newcommand{\mt}[1][]{\texttt{MT19937}\xspace}

\newcommand\citek[2][]{(see Knuth~\cite{#2}, #1)}

\title{Computationally Easy, Spectrally Good Multipliers for Congruential Pseudorandom Number Generators}

\author{Guy Steele \quad Sebastiano Vigna}

\begin{document}
\bibliographystyle{alpha}
\maketitle

\begin{abstract}
Congruential pseudorandom number generators rely on good \emph{multipliers}, that is, integers
that have good performance with respect to the spectral test.  We provide lists
of multipliers with a good lattice structure up to dimension eight and up to lag eight
for generators with typical power-of-two moduli, analyzing in detail
multipliers close to the square root of the modulus, whose product can be
computed quickly.
\end{abstract}

%

\section{Introduction}

A \emph{multiplicative congruential pseudorandom number generator} (MCG) is a computational process
defined by a recurrence of the form
\[
x_n = \bigl(a x_{n-1}\bigr)\bmod m,
\]
where $m\in\Z$ is the \emph{modulus}, $a\in\Z\cap[1\..m)$ is the \emph{multiplier}, and
$x_n\in\Z\cap[1\..m)$ is the \emph{state} of the generator after step $n$. Such
pseudorandom number generators (PRNGs) were introduced by
Lehmer~\cite{LehMMLSCU}, and have been extensively studied. If at each step we furthermore add a nonzero constant $c\in\Z\cap[1\..m)$, we obtain
a \emph{linear congruential pseudorandom number generator} (LCG), with state $x_n\in \Z\cap[0\..m)$:\footnote{We remark that these denominations, by now used for half a century, are completely wrong from a mathematical
viewpoint. The map $x\mapsto ax$ is indeed a \emph{linear} map, but the map $x\mapsto ax+c$ is an \emph{affine} map~\cite{BouEMA13}:
what we call an ``MCG'' or ``MLCG'' should called an ``LCG'' (this in fact happens in some books) 
and what we call an ``LCG'' should be called an ``ACG''. The mistake originated probably
in the interest of Lehmer in (truly) linear maps with prime moduli~\cite{LehMMLSCU}. Constants were added later to obtain large-period
generators with non-prime moduli, but the ``linear'' name stuck (albeit some authors are using the term ``mixed'' instead of ``linear''). At this point
it is unlikely that the now-traditional names will be corrected.}
\[
x_n = \bigl(a x_{n-1} + c\bigr)\bmod m.
\]
Under suitable conditions on $m$, $a$, and $c$, sequences of this
kind are periodic and their period is \emph{full}, that is, $m-1$ for MCGs ($c=0$) and $m$
for LCGs ($c\neq 0$). For MCGs, $m$ must be prime and $a$ must be a \emph{primitive element}
of the multiplicative group of residue classes $(\Z/m\Z)^\times$ (i.e., its powers must span the whole group). For LCGs, there are simple conditions that
must be satisfied~\citek[\S 3.2.1.2, Theorem A]{KnuACPII}.

For MCGs, when $m$ is not prime one can look for sequences that have \emph{maximum period}, that is, the largest possible period, given $m$.
We will be interested in moduli that are powers of two, in which case, if $m\geq 8$,
the maximum period is $m/4$, and the state must be odd~\citek[\S 3.2.1.2, Theorem B]{KnuACPII}.

While MCGs and LCGs have some known defects, they can be used in combination
with other pseudorandom number generators (PRNGs) or passed through
some output function that might lessen such
defects. Due to their speed and simplicity,
as well as a substantial accrued body of mathematical analysis,
they have been for a long time the PRNGs of choice in programming languages.

In this paper, we provide lists of multipliers for both MCGs and LCGs,
continuing the line of work by L'Ecuyer in his classic paper~\cite{LEcTLCGDSGLS}.
The quality of such multipliers is usually assessed by their score in the
\emph{spectral test}, described below.

The search for good multipliers is a sampling process from a large space: due to
the enormous increase in computational power in the last twenty years, we can now provide 
multipliers with significantly improved scores. In fact, for multipliers of up to $35$ bits we have now explored
the sample space exhaustively.

We consider only generators with power-of-two moduli; this choice
avoids the expensive modulo operation because nearly all contemporary
hardware supports binary arithmetic that is naturally carried modulo $2^w$ for some \emph{word size} $w$.
Such generators do have additional known, specific defects (e.g., the periods of the lowest bits are very short,
and the flip of a state bit will never propagate to lower bits), but there is a substantial body of literature
on how to reduce or avoid these defects. In particular,
it is now well understood that LCGs of this kind should not be used in isolation, but rather as components of composite (or combined) PRNGs.

Furthermore, in this paper we pay special attention to \emph{small} multipliers, that is, multipliers
close to the square root of the modulus $m$. For $m=2^{2w}$, this means
multipliers whose size in bits is $w\pm k$ for small $k$.  As is well known, many CPUs with natural
word size $w$ can produce with a single instruction, or two instructions, the full $2w$-bit product of two $w$-bit
operands, which makes such multipliers attractive from a computational viewpoint.

Unfortunately, such small  multipliers have known additional defects, which have been analyzed by
H\"{o}rmann and Derflinger~\cite{HoDPRNGWSRM}, who provided experimental
evidence of their undesirable behavior using a statistical test based on
rejection.

One of the goals of this paper is to deepen their analysis: while it is
known that $w$-bit multipliers for LCGs with power-of-two modulus $2^{2w}$
have inherent theoretical defects,  we show that these defects are reduced
as we add bits to the multiplier, and we quantify this improvement by
defining a new \emph{figure of merit} based on the magnitude on the
multiplier.  In the end, we provide tables of multipliers of $w+k$ bits,
where $k$ is relatively small, with quality closer to that of full
$2w$-bit multipliers.

The main motivation for our search is the implementation of 
a new family of splittable PRNGs that will be the core of the new
package for pseudorandom number generation in Java 17.~\cite{OOPSLA21}
The generators combine a LCG with power-of-2 modulus with an $\mathbf F_2$-linear
generator, and then apply a mixing function. In general, combination of PRNGs of different types was advocated already in 1984 by
Marsaglia,~\cite{MarCVRNG} and this particular combination has been popular
among programmers at least since 1990, as recently unearthed by a reverse engineering~\cite{LosSMW} of 
the well-known video game \emph{Super Mario World}. It has the useful property
of updating in parallel two components which potentially use different subunits of the CPU,
beside using the combination of two different kind of generators to improve quality; L'Ecuyer
and Granger--Pich\'e have studied in detail such combinations.\cite{LEGCGCDF}
In an effort to obtain a high-quality LCG component while minimizing CPU work, we were led
to the study of fast multipliers of good quality for the $128$-bit case, for which,
as we discuss in Section~\ref{sec:comp}, using (almost) half-width multipliers 
provides significant speed gains.

During the search for good multipliers, the authors have accumulated a large database
of candidates, which is publicly available for download, in case the reader is 
interested in looking for multipliers with specific properties. 
The software used to search for multipliers has been made public domain.
Both are available at \url{https://github.com/vigna/CPRNG}.

\section{Spectral figures of merit}

For every integer $d\geq 2$, the \emph{dimension}, we can consider the set of
$d$-dimensional points in the unit cube
\[
\Lambda_d = \Biggl\{\Biggl(\frac xm, \frac{f(x)}m, \frac{f^2(x)}m, \ldots, \frac{f^{d-1}(x)}m\Biggr) \Biggm| x\in\Z\cap[0\.. m)\Biggr\},
\]
where \[f(x)=(ax+c) \bmod m\] is the next-state map of a full-period generator.
This set is the intersection of a $d$-dimensional \emph{lattice} with the unit cube~\citek[\S 3.3.4.A]{KnuACPII}. Thus,
all points in $\Lambda_d$ lie on a family of equidistant, parallel \emph{hyperplanes}; in fact, there
are at most $(d!\,m)^{1/d}$ such hyperplanes~\cite{MarRNFMP}.   

The \emph{spectral test} (introduced by Coveyou and Macpherson~\cite{COMFAURNG}; see also Knuth~\cite{KnuACPII}, \S 3.3.4.G)
examines the family with the largest distance between
adjacent hyperplanes:
the smaller this \emph{largest interplane distance} is, the more evenly the generator fills the unit $d$-dimensional cube.
Using this idea, the \emph{figure of merit}
for dimension $d$ of an MCG or LCG is defined as
\[
f_d(m,a) =\frac{\nu_d}{\gamma_d^{1/2}\sqrtd{m}},
\]
where $1/\nu_d$ is the largest distance between adjacent hyperplanes found by considering all possible families of hyperplanes covering $\Lambda_d$.
We will usually imply the dependency on the choice of $m$ and $a$.

The definition of $f_d$ also relies on the \emph{Hermite constant} $\gamma_d$ for dimension $d$.  For $2\leq d \leq 8$, the Hermite constant has these values:
\[
\gamma_2=(4/3)^{1/2}, \gamma_3=2^{1/3}, \gamma_4=2^{1/2}, \gamma_5=2^{3/5}, \gamma_6=(64/3)^{1/6}, \gamma_7=4^{3/7}, \gamma_8=2.
\]
For all higher dimensions except $d=24$ only upper and lower bounds are known. 
Note that $1/\bigl(\gamma_d^{1/2}\sqrtd{m}\bigr)$ is
the smallest possible such largest interplane distance \citek[\S 3.3.4.E, equation (40)]{KnuACPII}; it follows that $0 < f_d\leq 1$.

\begin{figure}
\centering
\includegraphics{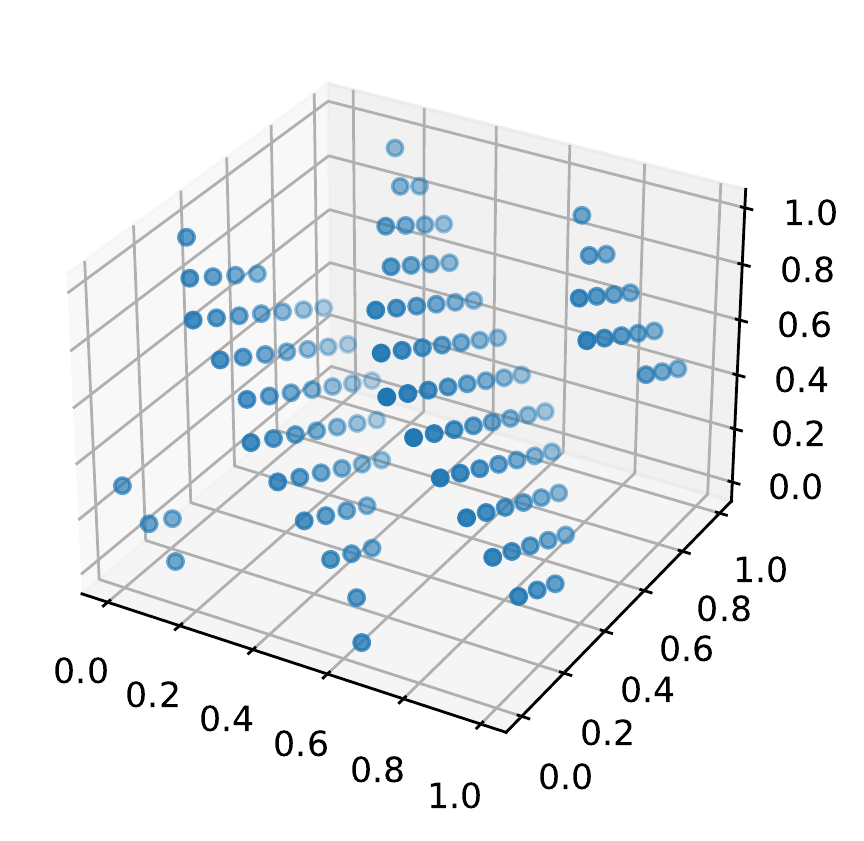}\includegraphics{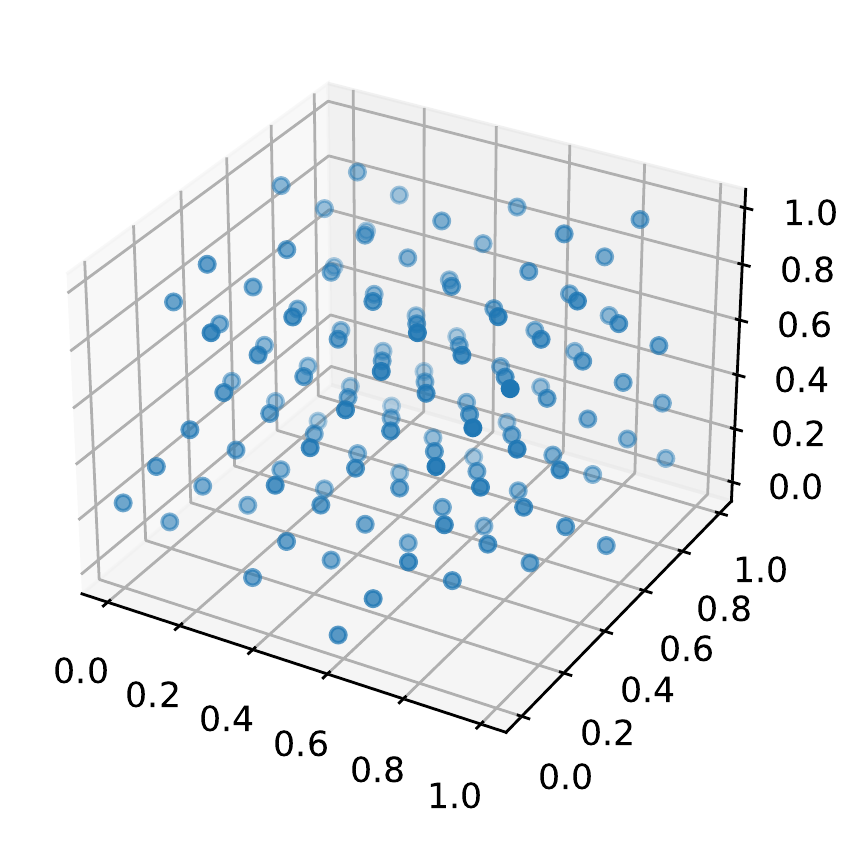}
\caption{\label{fig:hyper}A 3D example of the hyperplanes generated by two LCGs with $m=128$: $a=37$ (left, $f_3=0.545562$) or $a=29$ (right, $f_3=0.833359$).
Note the hyperplane structure of the multiplier with a low spectral score.}
\end{figure}

In Figure~\ref{fig:hyper} we plot in three dimensions $\Lambda_3$ for two LCGs with $m=128$. On the left,
we consider the multiplier $a=37$ with spectral score $f_3=0.545562$; on the right, the multiplier $a=29$ with spectral score $f_3=0.833359$.
It is evident how the $3$-dimensional points generated by the LCG with lower spectral score are arranged in a small number of hyperplanes,
and leave regions of unit cube empty, whereas the points generated by the LCG with higher spectral score fill the unit cube more evenly.

The reason for expressing the largest interplane distance in the form of a reciprocal $1/\nu_d$ is that $\nu_d$ is the \emph{length of the shortest
vector in the dual lattice $\Lambda^*_d$}. The dual lattice consists of all vectors whose scalar product with every vector
of the original lattice is an integer. In particular, it has the following basis~\citek[\S 3.3.4.C]{KnuACPII}:
\begin{align*}
(m, 0, 0, 0, & \ldots, 0, 0)\\
(-a, 1, 0, 0, &\ldots, 0, 0)\\
(-a^2, 0, 1, 0,& \ldots, 0, 0)\\
\vdots\;\, & \hphantom{\ldots,}\;\,\vdots \\  
(-a^{d-2}, 0, 0, 0,& \ldots, 1, 0)\\
(-a^{d-1}, 0, 0, 0,& \ldots, 0, 1).
\end{align*}
That is, $\Lambda^*_d$ is formed by taking all possible linear combinations of the vectors above with integer coefficients.
Note that the constant $c$ of an LCG has no role in the structure of $\Lambda_d$ and $\Lambda^*_d$, and that
we are under a full-period assumption.

The dual lattice is somewhat easier to work with, as its points have all integer coordinates; moreover, as we 
mentioned, if we call $\nu_d$ the length of its shortest vector, the maximum distance between parallel hyperplanes
covering $\Lambda_d$ is $1/\nu_d$ (and, indeed, this is how the figure of merit $f_d$ is computed).

A deeper analysis can be obtained by applying the spectral test to \emph{sequences with lag $\ell$},~\cite{LEcBLSVNVPSLR} that is,
lattices generated by sequences ${x}$,~$f^\ell(x)$,~$f^{2\ell}(x)$, $\dots$.
We will use the characterizations proved by Entacher~\cite{EntPSLRNST} to apply the spectral test to such sequences, too.

\section{Computationally easy multipliers}
\label{sec:comp}

Multipliers smaller than $\sqrt m$ have been advocated~\cite{DoRMCPSRG,LGMPRNGS}, in particular when the modulus is a power of two, say $m=2^{2w}$, because they
do not require a full $2w$-bit multiplication: writing $x_-$ and $x^-$ for the $w$ lowest and highest bits,
respectively, of a $2w$-bit value $x$ (that is,  $x_-=x \bmod 2^w$ and $x^-=\lfloor x/2^w\rfloor$), we have 
\[
(ax)\bmod 2^{2w} = \bigl(ax_- + a\cdot 2^w x^- \bigl)\bmod 2^{2w} = \bigl(a x_- + 2^w\cdot a x^- \bigl) \bmod 2^{2w}.
\]
The first multiplication, $ax_-$, has a $2w$-bit operand $a$ and a $w$-bit operand $x_-$, and in general the result may be $2w$ bits wide;
but the second multiplication, $ax^-$, can
be performed by an instruction that takes two $w$-bit operands and produces only a $w$-bit result that is only the low $w$ bits of the full product,
because the modulo operation effectively discards the high $w$ bits of that product.
Moreover, if the multiplier $a=2^w a^-+a_-$ has a high part that is small (say, $a^- < 256$) or of a special form (for example, $a^- = j2^n$ where $j$ is $1$, $3$, $5$, or $9$), then the first multiplication may also be computed using a faster method.
Contemporary optimizing compilers know how to exploit such special cases, perhaps by using a small immediate operand rather than loading the entire multiplier into
a register, or perhaps by using shift instructions and/or such instructions as \texttt{lea} (Load Effective Address), which in the Intel 64-bit architecture may be used to compute $x+jy$ on two 64-bit operands $x$ and $y$ for $j$ = $2$, $4$, or $8$\cite{INTEL-INSTRUCTION-SET}.
And even if the compiler produces the same code for, say, a multiplier that is $(3/2)w$ bits wide as for a multiplier that is $2w$ bits wide,
some hardware architectures may notice the smaller multiplier on the fly and handle it in a faster way.

Multiplication by a constant $a$ of size $w$, that is, of the form $a_-$ (in other words, $a^-=0$), is especially simple:
\[
a_-x \bmod 2^{2w}  = \bigl( a_-\,x_- + 2^w a_-\, x^- \bigr)\bmod 2^{2w} \hfil
\]
Notice that the addition can be performed as a $w$-bit addition of the low $w$ bits of $a_- x^-$ into the high half of $a_-\, x_-$.

In comparison, multiplication by a constant $a$ of size $w+1$, that is, of the form $2^w+a_-$ (in other words, $a^-=1$), requires only one extra addition:
\begin{multline*}
\bigl(\bigl(2^w + a_-\bigr)x\bigr)  \bmod 2^{2w}  = \bigl(\bigl(2^w + a_-\bigr)x_- + \bigl(2^w + a_-\bigr)\bigl(x^- \cdot 2^w\bigr)\bigr) \bmod 2^{2w}  =\\
\bigl( 2^wx_- + a_-\, x_- + 2^w\cdot a_-\, x^- \bigr)\bmod 2^{2w} = 
\bigl( a_-\, x_- + 2^w\cdot (x_- + a_-\, x^- )\bigr)\bmod 2^{2w}. 
 \end{multline*}
Modern compilers know the reduction above and will reduce the strength of operations involved as necessary.

Even without the help of the compiler, we can push this idea further to multipliers of the form $2^{w+k}+ a$, where $k$ is a small positive integer constant:
\begin{multline*}
\bigl(\bigl(2^{w+k} + a\bigr)x\bigr)  \bmod 2^{2w}  = \bigl(\bigl(2^{w+k} + a\bigr)x_- + \bigl(2^{w+k} + a\bigr)\bigl(x^- \cdot 2^w\bigr)\bigr) \bmod 2^{2w}  =\\
\bigl(  2^{w+k}x_- + a x_- +  2^w\cdot a x^- \bigr)\bmod 2^{2w} = 
\bigl( a x_- + 2^w\cdot \bigl(2^k x_- + a x^- \bigr)\bigr)\bmod 2^{2w}. 
 \end{multline*}
In comparison to the $(w+1)$-bit case, we just need an additional shift to compute $2^k x_-$.
In the interest of efficiency, it thus seems interesting to study in more detail
the quality of small multipliers.

In Figure~\ref{fig:intelcode}
we show code generated by the \texttt{clang} compiler
that uses 64-bit instructions to multiply a 128-bit value (in registers \texttt{rsi} and \texttt{rdi})
by (whimsically chosen) constants of various sizes.
The first example shows that if the constant is of size 64, indeed only two 64-bit by 64-bit multiply instructions
(one producing a 128-bit result and the other just a 64-bit result) and one 64-bit add instruction are needed.
The second example shows that if the constant is of size 65, indeed only one extra 64-bit add instruction is needed.
For constants of size 66 and above, more
sophisticated strategies emerge that use \texttt{leaq} (the quadword, that is, 64-bit form of \texttt{lea}) and shift instructions and even subtraction. In Figure~\ref{fig:armcode}
we show three examples of code generated by \texttt{clang} for the ARM processor: since its RISC architecture~\cite{ARM-INSTRUCTION-SET}
can only load constant values $16$ bits at a time, the length of the sequence of instructions grows
as the multiplier size grows.  On the other hand, note that the ARM architecture has a multiply-add instruction \texttt{madd}.

\begin{figure}[p]
\centering
\begin{tabular}{@{}rrc@{}}
Bits&\multicolumn{1}{@{}c}{Multiplier} & Code \\[1ex]
\hline
64&\texttt{0xCAFEF00DDEADF00D} &\begin{minipage}{5cm}\footnotesize%
\vspace*{0.4em}
\begin{verbatim}
movabsq $0xCAFEF00DDEADF00D, %rax
imulq   %rax, %rsi
mulq    %rdi
addq    %rsi, %rdx
\end{verbatim}%
\vspace*{0.2em}
\end{minipage}\\
\hline
65&\texttt{0x1CAFEF00DDEADF00D} &
\begin{minipage}{5cm}\footnotesize%
\vspace*{0.4em}
\begin{verbatim}
movabsq $0xCAFEF00DDEADF00D, %rcx
imulq   %rcx, %rsi
mulq    %rcx
addq    %rdi, %rdx
addq    %rsi, %rdx
\end{verbatim}
\vspace*{0.2em}
\end{minipage}\\
\hline
66&\texttt{0x2CAFEF00DDEADF00D}& \begin{minipage}{5cm}\footnotesize%
\vspace*{0.4em}
\begin{verbatim}
movabsq $xCAFEF00DDEADF00D, %rcx
imulq   %rcx, %rsi
mulq    %rcx
leaq    (%rdx,%rdi,2), %rdx
addq    %rsi, %rdx
\end{verbatim}%
\vspace*{0.2em}
\end{minipage}\\
%
\hline
67&\texttt{0x4CAFEF00DDEADF00D} &
\begin{minipage}{5cm}\footnotesize%
\vspace*{0.4em}
\begin{verbatim}
movabsq $0xCAFEF00DDEADF00D, %rcx
imulq   %rcx, %rsi
mulq    %rcx
leaq    (%rdx,%rdi,4), %rdx
addq    %rsi, %rdx
\end{verbatim}%
\vspace*{0.2em}
\end{minipage}\\
\hline
67&\texttt{0x5CAFEF00DDEADF00D} &
\begin{minipage}{5cm}\footnotesize%
\vspace*{0.4em}
\begin{verbatim}
movabsq $0xCAFEF00DDEADF00D, %rcx
mulq    %rcx
imulq   %rcx, %rsi
leaq    (%rdi,%rdi,4), %rcx
addq    %rcx, %rdx
addq    %rsi, %rdx
\end{verbatim}%
\vspace*{0.2em}
\end{minipage}\\
%
\hline
67&\texttt{0x7CAFEF00DDEADF00D} &
\begin{minipage}{5cm}\footnotesize%
\vspace*{0.4em}
\begin{verbatim}
movabsq $0xCAFEF00DDEADF00D, %r8
mulq    %r8
leaq    (,%rdi,8), %rcx
subq    %rdi, %rcx
addq    %rcx, %rdx
imulq   %r8, %rsi
addq    %rsi, %rdx
\end{verbatim}%
\vspace*{0.2em}
\vspace*{0.2em}
\end{minipage}\\
\hline
96&\texttt{0xFADC0C0ACAFEF00DDEADF00D} &
\begin{minipage}{5cm}\footnotesize%
\vspace*{0.4em}
\begin{verbatim}
movl    $0xFADC0C0A, %ecx
movabsq $0xCAFEF00DDEADF00D, %r8
mulq    %r8
imulq   %rdi, %rcx
addq    %rcx, %rdx
imulq   %r8, %rsi
addq    %rsi, %rdx
\end{verbatim}%
\vspace*{0.2em}
\end{minipage}\\
\hline
128&\texttt{0xAB0DE0FBADC0FFEECAFEF00DDEADF00D} &
\begin{minipage}{5cm}\footnotesize%
\vspace*{0.4em}
\begin{verbatim}
movabsq $0xAB0DE0FBADC0FFEE, %rcx
movabsq $0xCAFEF00DDEADF00D, %r8
mulq    %r8
imulq   %rdi, %rcx
addq    %rcx, %rdx
imulq   %r8, %rsi
addq    %rsi, %rdx
\end{verbatim}%
\vspace*{0.2em}
\end{minipage}
\end{tabular}
\caption{\label{fig:intelcode}\texttt{clang}-generated Intel code for the multiplication part of a $128$-bit LCG using
multipliers of increasing size. The code generated for more than $96$ bits (not shown here) is identical to the $128$-bit case.}
\end{figure}

\begin{figure}[p]
\centering
\begin{tabular}{@{}rrc@{}}
Bits&\multicolumn{1}{@{}c}{Multiplier} & Code \\[1ex]
\hline
64&\texttt{0xCAFEF00DDEADF00D} &\begin{minipage}{5cm}\footnotesize%
\vspace*{0.4em}
\begin{verbatim}
mov     x8, #0xF00D
movk    x8, #0xDEAD, lsl #16
movk    x8, #0xF00D, lsl #32
movk    x8, #0xCAFE, lsl #48
umulh   x9, x0, x8
madd    x1, x1, x8, x9
mul     x0, x0, x8
\end{verbatim}%
\vspace*{0.2em}
\end{minipage}\\
\hline
65&\texttt{0x1CAFEF00DDEADF00D} &
\begin{minipage}{5cm}\footnotesize%
\vspace*{0.4em}
\begin{verbatim}
mov     x8, #0xF00D
movk    x8, #0xDEAD, lsl #16
movk    x8, #0xF00D, lsl #32
movk    x8, #0xCAFE, lsl #48
umulh   x9, x0, x8
add     x9, x9, x0
madd    x1, x1, x8, x9
mul     x0, x0, x8
\end{verbatim}
\vspace*{0.2em}
\end{minipage}\\
\hline
67&\texttt{0x7CAFEF00DDEADF00D} &
\begin{minipage}{5cm}\footnotesize%
\vspace*{0.4em}
\begin{verbatim}
mov     x8, #0xF00D
movk    x8, #0xDEAD, lsl #16
movk    x8, #0xF00D, lsl #32
movk    x8, #0xCAFE, lsl #48
lsl     x9, x0, #3
umulh   x10, x0, x8
sub     x9, x9, x0
add     x9, x10, x9
madd    x1, x1, x8, x9
mul     x0, x0, x8
\end{verbatim}%
\vspace*{0.2em}
\vspace*{0.2em}
\end{minipage}\\
\hline
96&\texttt{0xFADC0C0ACAFEF00DDEADF00D} &
\begin{minipage}{5cm}\footnotesize%
\vspace*{0.4em}
\begin{verbatim}
mov     x8, #0xF00D
movk    x8, #0xDEAD, lsl #16
movk    x8, #0xF00D, lsl #32
movk    x8, #0xCAFE, lsl #48
mov     w9, #0x0C0A
movk    w9, #0xFADC, lsl #16
umulh   x10, x0, x8
madd    x9, x0, x9, x10
madd    x1, x1, x8, x9
mul     x0, x0, x8
\end{verbatim}%
\vspace*{0.2em}
\end{minipage}\\
\hline
128&\texttt{0xAB0DE0FBADC0FFEECAFEF00DDEADF00D} &
\begin{minipage}{5cm}\footnotesize%
\vspace*{0.4em}
\begin{verbatim}
mov     x9, #0xF00D
mov     x8, #0xFFEE
movk    x9, #0xDEAD, lsl #16
movk    x8, #0xADC0, lsl #16
movk    x9, #0xF00D, lsl #32
movk    x8, #0xE0FB, lsl #32
movk    x9, #0xCAFE, lsl #48
movk    x8, #0xAB0D, lsl #48
umulh   x10, x0, x9
madd    x8, x0, x8, x10
madd    x1, x1, x9, x8
mul     x0, x0, x9
\end{verbatim}%
\vspace*{0.2em}
\end{minipage}
\end{tabular}
\caption{\label{fig:armcode}\texttt{clang}-generated ARM code for the multiplication part of a $128$-bit LCG using
multipliers of increasing size. Note how the number of \texttt{mov} and \texttt{movk} instructions depends on the size of the multiplier.}
\end{figure}

To verify whether the reduction in strength and code size has an impact
on the speed of the generator, we have run a number of microbenchmarks
for the $128$-bit case (our main motivation) on
an Intel\textregistered{} Core\texttrademark{} i7-8700B CPU @$3.20$\,GHz (Haswell)
and on an AWS Graviton 2 processor based on $64$-bit Arm Neoverse cores @$2.5$\,GHz.
We provide results for both \texttt{gcc} 10.2.1 and \texttt{clang} 10.0.1.
We disabled loop unrolling (to avoid different unrolling strategies for
different functions or by different compilers) and inlining (because it is the only
safe way to guarantee that we are actually measuring the whole next-state function,
including loading of constants). We used the CPU governance tools, where available,
to stabilize the clock speed. 

Table~\ref{tab:benchmarks} contains the results of
the benchmarks for the smallest and largest interesting sizes, showing clearly
the advantage of small multipliers. Intermediate results (not shown in the table)
confirm the trend from faster to slower, albeit sometimes the scheduler
of one compiler makes a bad choice for a particular size and architecture.
Note that the timings 
include the looping time and the function call time: thus, relative
timing differences between the actual next-state functions are larger
than the relative differences between the displayed values.

We remark that different architectures might give different results, and that
different compilers (and even different releases of the same compiler) have different scheduling models.
Thus, while smaller multipliers generally tend to allow a compiler to emit shorter and possibly weaker instructions,
rerunning our benchmarks in a different setup might thus give different results.
The best practice is to benchmark a PRNG in the context of the application
that consumes its output.

\begin{table}
\begin{center}
\renewcommand{\arraystretch}{1.2}
\begin{tabular}{lrrrr}
\hline
\multicolumn{5}{c}{LCG}\\
\hline
&\multicolumn{2}{c}{gcc}&\multicolumn{2}{c}{clang}\\
\hline
& i7 &  ARM & i7 & ARM  \\
\hline
$64$ bits  &  1.201 &  4.001 &  1.201 &  4.001\\
$65$ bits  &  1.441 &  4.001 &  1.306 &  4.001\\
$128$ bits  &  1.680 &  5.201 &  1.444 &  5.201\\
\hline
\multicolumn{5}{c}{MCG}\\
\hline
&\multicolumn{2}{c}{gcc}&\multicolumn{2}{c}{clang}\\
\hline
& i7  & ARM & i7 & ARM  \\
\hline
$64$ bits  &  1.202 &  4.001 &  1.273 &  4.001\\
$128$ bits  &  1.299 &  5.201 &  1.438 &  5.201\\
\hline
\end{tabular}
\end{center}
\caption{\label{tab:benchmarks}Microbenchmarks on different architectures and compilers for the smallest
and largest multiplier sizes in the case of $128$ bits of state.
Timings are in nanosecond per iteration of the next-state function. Relative standard deviation on 10 repeats is below $1\%$.}
\end{table}

\section{Bounds on spectral scores}

If the multiplier is smaller than the root of order $d$ of the modulus, we can 
compute exactly the figure of merit $f_d$:
\begin{proposition}
\label{prop:bound}
Consider a full-period LCG with modulus $m$ and multiplier $a$.
Then, for every $d\geq2$, if $a<\sqrtd{m}$ we have $\nu_d=\sqrt{a^2+1}$,
and it follows that
\[
f_d = \frac{\sqrt{a^2+1}}{\gamma_d^{1/2}\sqrtd{m}}
\]
\end{proposition}
A slightly weaker result with just an upper bound for $f_d$ is reported in~\cite{LEcBLSVNVPSLR}.

\begin{proof}
The length $\nu_d$ of the shortest vector of the dual lattice $\Lambda^*_d$ can be
easily written as 
\begin{equation}
\label{eq:vec}
\nu_d =\min_{(x_0,\ldots,x_{d-1})\neq (0,\ldots,0)}  \Bigl\{\sqrt{x_0^2+x_1^2+\cdots+x_{d-1}^2} \Bigm| x_0+a x_1 + a^2 x_2 + \cdots + a^{d-1}x_{d-1} \equiv 0 \mod m\Bigr\},
\end{equation}
where $(x_0,\ldots,x_{d-1})\in \Z^d$, due to the simple structure of the basis of $\Lambda^*_d$~\citek[\S 3.3.4]{KnuACPII}.
Clearly, in general $\nu_d\leq\sqrt{a^2+1}$, because $(-a,1,0,0,\ldots,0)\in\Lambda^*_d$.\footnote{This is what was shown in~\cite{LEcBLSVNVPSLR}, without the condition that $a<\sqrtd{m}$.}
However, when $a<\sqrtd{m}$ we have $\nu_d = \sqrt{a^2+1}$, because no vector shorter than $\sqrt{a^2+1}$ 
can fulfill the modular condition.

To prove this statement, note that a vector $(x_0,\ldots,x_{d-1})\in \Lambda^*_d$ shorter than $\sqrt{a^2+1}$ must have all coordinates smaller than 
$a$ in absolute value
(if one coordinate has absolute value $a$, all other coordinates must be zero, or the vector would have length at least $\sqrt{a^2+1}$, so the vector cannot belong to $\Lambda_d^*$).
Then, for every $0\leq j < d$ 
\[\left|\sum_{i=0}^{j} x_i a^i\right|\leq \sum_{i=0}^{j} \bigl|x_i\bigr| a^j < a^{j+1} < m,\]
so the modular condition in~(\ref{eq:vec}) must be fulfilled by equality with zero. However, let 
$t$ be the index of the last nonzero component of $(x_0,\ldots,x_{d-1})$ (i.e., $x_i=0$ for $i>t$):  
then, $\bigl|\sum_{i=0}^{t-1} x_i a^i\bigr|< a^t,$ whereas $|x_ta^t|\geq a^t$, so their sum cannot be zero.
\end{proof}
Note that if $m=a^d$, then the vector that is $a$ in position $d-1$ and zero elsewhere is in $\Lambda_d^*$, but by the proof
above shorter vectors cannot be, so
\[
f_d = \frac{a}{\gamma_d^{1/2}\sqrtd{m}}=\frac1{\gamma_d^{1/2}}.
\]

Using the approximation $\sqrt{a^2+1}\approx a$, this means that if $a\leq\sqrtd{m}$ then for $2\leq d \leq 8$, $f_d$ cannot be greater than approximately
\begin{multline*}
(4/3)^{-1/4}\approx 0.9306,\quad 2^{-1/6}\approx 0.8909,\quad 2^{-1/4}\approx 0.8409,\quad 2^{-6/10} \approx 0.8122,\quad \\(64/3)^{-1/12}\approx 0.7749,\quad 4^{-3/14}\approx 0.7430,\quad 2^{-1/2}\approx 0.7071
\end{multline*}
for $d=2,\ldots, 8$.
For $d>2$ this is not a problem, as such very small multipliers are not commonly used. However, choosing a multiplier that is smaller than or
equal to $\sqrt m$ has the effect of making it impossible to obtain a figure of merit close to $1$ in dimension $2$.
%
Note that, for any $d$, as $a$ drops well below $\sqrtd{m}$ the figure of merit $f_d$ degenerates quickly;
for example, if $a<\sqrt m / 2$ then $f_2$ cannot be greater than $(4/3)^{-1/4} / 2 \approx 0.4653$ (the
fact that multipliers much smaller than $\sqrt m$ were unsatisfactory was noted informally already by Downham and Roberts~\cite{DoRMCPSRG}).

Nonetheless, as soon
as we allow $a$ to be even a tiny bit larger than $\sqrt{m}$, $\nu_2$ (and thus $f_2$) is no longer constrained: indeed, if $m=2^{2w}$, a $(w+1)$-bit multiplier
is sufficient to get a figure of merit in dimension $2$ very close to $1$ (larger than $0.998$); see Table~\ref{tab:lcgcomp}.

MCGs with power-of-two moduli cannot achieve full period: the maximum period is $m/4$. 
It turns out that the lattice structure, however, is very similar to the full-period case, once
we replace $m$ with $m/4$ in the definition of the dual lattice.
Correspondingly, we have to replace $\sqrtd{m}$ with $\sqrtd{m/4}$~\citek[\S 3.3.4, Exercise 20]{KnuACPII}:
\begin{proposition}
\label{prop:bound2}
Consider an MCG with power-of-two modulus $m$, multiplier $a$, and period $m/4$.
Then for every $d\geq2$ and every $a<\sqrtd{m/4}$ we have $\nu_d=\sqrt{a^2+1}$,
and it follows that
\[
f_d = \frac{\sqrt{a^2+1}}{\gamma_d^{1/2}\sqrtd{m/4}}.
\]
\end{proposition}
Note that Proposition~\ref{prop:bound2} imposes limits on the figures of merit for
$(w-1)$-bit multipliers for $2w$-bit MCGs, but does not impose any limits
on $w$-bit multipliers for $2w$-bit MCGs.
In Table~\ref{tab:mcgcomp}, observe that the 31-bit multipliers necessarily have figures of merit $f_2$ smaller than
$(4/3)^{-1/4}\approx 0.9306$ (though one value for $f_2$, namely $0.930577$, is quite close),
but for multipliers of size $32$ and greater we have been able to choose examples
for which $f_2$ is well above $0.99$.

\section{Beyond spectral scores}

Proposition~\ref{prop:bound} does not impose a bound on $(w+1)$-bit multipliers,
and indeed, as we remarked in the previous section, one can find $(w+1)$-bit multipliers
whose spectral scores are similar to those of
$2w$-bit multipliers. We now show that, however, on closer inspection, the spectral
scores are not telling the whole story.

H\"{o}rmann and Derflinger~\cite{HoDPRNGWSRM} studied multipliers close to the square root of the modulus for LCGs
with $32$ bits of state,
and devised a statistical test that makes generators using such multipliers fail: the intuition behind the test
is that with such multipliers there is a relatively short lattice vector $\bm s = (1/m, a/m)\in \Lambda_2$ that is almost parallel to the $y$ axis. The existence of this vector
creates bias in pairs of consecutive outputs, a bias that can be detected by generating random variates from a distribution
using the rejection method: if at some point the density of the distribution increases sharply,
the rejection method will underrepresent certain parts of the distribution and overrepresent others.

We applied an instance of the H\"{o}rmann--Derflinger test to congruential generators (both LCG and MCG) with
$64$ bits of state using a Cauchy distribution on the interval $[-2\..2)$.
We divide the interval into $10^8$ slots that contain the same probability mass,
repeatedly generate by rejection $10^9$ samples from the distribution, and compute a
$p$-value using a $\chi^2$ test on the slots. We consider the number of repetitions required to bring
the $p$-value is very close to zero\footnote{More precisely, when the $p$-value returned
by the Boost library implementation of the $\chi^2$ test becomes zero, which in this case
happens when the $p$-value goes below $\approx 10^{-16}$.} a measure of the
resilience of the multiplier to the \HD test, and thus a positive feature (that is, a larger number is better).

The results are reported in the last column of Tables~\ref{tab:lcgcomp} and~\ref{tab:mcgcomp}. 
As we move from small to large multipliers, the number of iterations necessary to detect
bias grows, but among multipliers with the same number of bits there is a very large
variability.\footnote{We also tested a generator with $128$ bits of state and a $64$-bit multiplier, but at that size the bias
is undetectable even with a hundred times as many ($10^{10}$) slots.}


The marked differences have a simple explanation: incrementing the
number of bits does not translate immediately into a significantly longer vector $\bm s$.
To isolate generators in which $\bm s$ is less pathological, we have to consider larger multipliers,
as $\|\bm s \|=\sqrt{a^2+1}/m$. In particular, we define the 
simple figure of merit $\lambda$ for a full-period LCG as
\[
\lambda = \frac{\|\bm s\|}{1 / \sqrt m}=\frac{\sqrt{a^2+1}/m}{1 / \sqrt m} = \frac{\sqrt{a^2+1}}{\sqrt m}\approx a / \sqrt m
\]
In other words, we measure the length of $\bm s$ with respect to the threshold $1/\sqrt m$ of Proposition~\ref{prop:bound}.
In general, for a set of multipliers bounded by $B$, $\lambda \leq B/\sqrt m$.

Note that because of Proposition~\ref{prop:bound}, if $a < \sqrt m$
\[
f_2 / \lambda =   \frac{\sqrt{a^2+1}}{\gamma_2^{1/2}\sqrt m}\big\slash \frac{\sqrt{a^2+1}}{\sqrt m}= \gamma_2^{-1/2}\approx 0.9306,
\]
that is, for multipliers smaller than $\sqrt m$ the two figures of merit $f_2$ and $\lambda$ are linearly dependent. Just one additional
bit, however, makes the two figures of merit $f_2$ and $\lambda$ no longer linearly dependent (see the entries for $33$-bit multipliers in Table~\ref{tab:lcgcomp},
as well as the entries for $32$-bit multipliers in Table~\ref{tab:mcgcomp}). 

For MCGs with power-of-two modulus $m$, $\bm s = (4/m, 4a/m)$, and, in view of Proposition~\ref{prop:bound2}, we define 
\[
\lambda = \frac{\|\bm s\|}{1 / \sqrt{m/4}}=\frac{\sqrt{a^2+1}/(m/4)}{1 / \sqrt{m/4}} = \frac{\sqrt{a^2+1}}{\sqrt {m/4}}\approx 2 a / \sqrt m
\]

\begin{table}
\centering
\renewcommand{\arraystretch}{1.4}
\setlength{\tabcolsep}{8pt}
\begin{tabular}{lrrrr}
\multicolumn{1}{c}{Bits}&\multicolumn{1}{c}{$a$}&\multicolumn{1}{c}{$f_2$}&\multicolumn{1}{c}{$\lambda$}&\multicolumn{1}{c}{H--D}\\\hline
\multirow{2}{*}{$32$}
	&	\texttt{0xfffeb28d}	&	$0.930586$	&	$\phantom{00}1.00$	&	$6$\\
	&	\texttt{0xcffef595}	&	$0.756102$	&	$\phantom{00}0.81$	&	$4$\\
\hline\multirow{2}{*}{$33$}
	&	\texttt{0x1dd23bba5}	&	$0.998598$	&	$\phantom{00}1.86$	&	$19$\\
	&	\texttt{0x112a563ed}	&	$0.998387$	&	$\phantom{00}1.07$	&	$7$\\
\hline\multirow{2}{*}{$34$}
	&	\texttt{0x3de4f039d}	&	$0.998150$	&	$\phantom{00}3.87$	&	$72$\\
	&	\texttt{0x2cfe81d9d}	&	$0.992874$	&	$\phantom{00}2.81$	&	$46$\\
\hline\multirow{2}{*}{$35$}
	&	\texttt{0x78ad72365}	&	$0.995400$	&	$\phantom{00}7.54$	&	$313$\\
	&	\texttt{0x49ffd0d25}	&	$0.991167$	&	$\phantom{00}4.62$	&	$109$\\

\end{tabular}
\caption{\label{tab:lcgcomp}A comparison of small LCG multipliers for $m=2^{64}$.
In the $32$-bit case, $f_2$ and $\lambda$ are linearly dependent,
and $f_2$ is necessarily smaller than approximately $0.9306$.
For sizes above $32$ bits we show multipliers
all with almost perfect $f_2$ (well above $0.99$) but different $\lambda$.
The last column shows the corresponding number of iterations of the \HD test.}
\end{table}

\begin{table}
\centering
\renewcommand{\arraystretch}{1.4}
\setlength{\tabcolsep}{8pt}
\begin{tabular}{lrrrr}
\multicolumn{1}{c}{Bits}&\multicolumn{1}{c}{$a$}&\multicolumn{1}{c}{$f_2$}&\multicolumn{1}{c}{$\lambda$}&\multicolumn{1}{c}{H--D}\\\hline
\multirow{2}{*}{$31$}
	&	\texttt{0x7ffc9ef5}	&	$0.930509$	&	$\phantom{00}0.50$	&	$2$\\
	&	\texttt{0x672a3fb5}	&	$0.750046$	&	$\phantom{00}0.40$	&	$1$\\
\hline\multirow{2}{*}{$32$}
	&	\texttt{0xef912f85}	&	$0.994558$	&	$\phantom{00}0.94$	&	$4$\\
	&	\texttt{0x89f353b5}	&	$0.997577$	&	$\phantom{00}0.54$	&	$2$\\
\hline\multirow{2}{*}{$33$}
	&	\texttt{0x1f0b2b035}	&	$0.996853$	&	$\phantom{00}1.94$	&	$22$\\
	&	\texttt{0x16aa7d615}	&	$0.994427$	&	$\phantom{00}1.42$	&	$11$\\
\hline\multirow{2}{*}{$34$}
	&	\texttt{0x3c4b7aba5}	&	$0.992314$	&	$\phantom{00}3.77$	&	$81$\\
	&	\texttt{0x2778c3815}	&	$0.998339$	&	$\phantom{00}2.47$	&	$37$\\
\hline\multirow{2}{*}{$35$}
	&	\texttt{0x7d3f85c05}	&	$0.998470$	&	$\phantom{00}7.83$	&	$354$\\
	&	\texttt{0x40dde345d}	&	$0.996172$	&	$\phantom{00}4.05$	&	$87$\\

\end{tabular}
\caption{\label{tab:mcgcomp}A comparison of small MCG multipliers for $m=2^{64}$.
In the $31$-bit case, $f_2$ and $\lambda$ are linearly dependent,
and $f_2$ is necessarily smaller than approximately $0.9306$.
For each size above 31 we show multipliers
with almost perfect $f_2$ but different $\lambda$.
The last column shows the corresponding number of iterations of the \HD test.}
\end{table}

In Tables~\ref{tab:lcgcomp} and~\ref{tab:mcgcomp} we report a few small-sized
multipliers together with the figures of merit
$f_2$ and $\lambda$, as well as the number of iterations required by our use of the \HD test: larger values
of $\lambda$ (i.e., larger multipliers) correspond to more resilience to the test.

\section{Potency}

\emph{Potency} is a property of multipliers of LCGs: it is defined as the
minimum $s$ such that $(a-1)^s$ is a multiple of the modulus $m$. Such an $s$
always exists for full-period multipliers, because one of the conditions for
full period is that $a-1$ be divisible by every prime that divides $m$ (when
$m$ is a power of two, this simply means that $a$ must be odd).

Multipliers of low potency generate sequences that do not look very random: in the
case $m$ is a power of two, this is very immediate, as a multiplier $a$ with low
potency is such that $a-1$ is divisible by a large power of two, say, $2^k$. In this
case, the $k$ lowest bits of $ax$ are the same as the $k$ lowest bits of $x$, which means that changes to
the $k$ lowest bits of the state depend only on the fact that we add $c$. For this
reason, one ordinarily chooses multipliers of maximum possible potency,\footnote{Note that ``maximum possible potency'' is a quite
rough statement, because potency is a very rough measure when applied to multipliers that are powers of primes: for example,
when $m=2^{2w}$ a generator with $a-1$ divisible by $2^w$ (but not by $2^{w+1}$) and a generator with $a-1$ divisible
by $2^{2w-1}$ have both potency $2$, but in view of the discussion above their randomness is very different. More precisely, here we choose
to consider only multipliers which leave unchanged that smallest possible number of lower bits.} and since for full period if $m$ is a multiple
of four, then $a-1$ must be a multiple of four, we have to choose $a$ so that $(a-1)/4$ is odd,
that is, $a\bmod 8 = 5$.

Potency has an interesting interaction with the constant $c$, described for the first time
by Durst~\cite{DurULCGPRNG} in response to proposals from Percus and Kalos~\cite{PeKRNGMPP} and Halton~\cite{HalPRT}
to use different constants to generate different streams for multiple processors. If we
take a multiplier $a$ and a constant $c$, then for every $r\in\Z/m\Z$ the generator with
multiplier $a$ and constant $(a-1)r +c$ has the same sequence of the first one, up to 
addition with $r$. Indeed, if we consider sequences starting from $x_0$ and $y_0 = x_0 - r$, we have\footnote{All remaining computations in this section are performed in $\Z/m\Z$.} 
\[
y_n = a y_{n-1} + (a-1)r + c = a (x_{n-1} - r) + (a-1)r + c  = x_n - r.  
\]
That is, for a fixed multiplier $a$, the constants $c$ are divided into classes by the equivalence relation of generating
the same sequence up to an additive constant.

How many classes do exist? The answer depends on the potency of $a$, as it comes down to solving
the modular equation
\[
c' - c =(a-1)r 
\]
If $a$ has low potency, this equation will be rarely solvable because there will be many equivalence classes: but for the specific case
where $m$ is a power of two and $a\bmod 8 = 5$, it turns out that there are just two classes: the
class of constants that are congruent to $1$ modulo $4$, and the class of constants that
are congruent to $3$ modulo $4$. All constants in the first class yield the sequence
$x_n = ax_{n-1}+1$, up to an additive constant, and all constants in the second
class yield the sequence $x_n = ax_{n-1}-1$, up to an additive constant.
It follows that if one tries to use three (or more) different streams, even if one chooses different
constants for the streams, at least two of the streams will be correlated.

If we are willing to weaken slightly our notion of equivalence, in this case
we can extend Durst's considerations: if we consider sequences starting from $x_0$ and $y_0 = - x_0 + r$, then 
\[
y_n = a y_{n-1} - ((a-1)r + c) = a (-x_{n-1} + r) - (a-1)r - c  = -x_n + r.  
\]
Thus, if we consider the equivalence relation of generating sequences that are the same up to an additive constant \emph{and} possibly a sign change,
then all sequences generated by a multiplier $a$ of 
maximum potency for a power-of-two modulus $m$ are the same, because to prove equivalence we now need to solve just
one of the two modular equations
\[
c' - c =(a-1)r \qquad \hbox{and} \qquad c' + c = (a-1)r, 
\]
and while the first equation is solvable when the residues of $c$ and $c'$ modulo $4$ are the
same, the second equation is solvable when the residues are different.

All in all, if one strives for a generator of higher quality (i.e., maximum potency), as we will do in Section~\ref{sec:search}, changing
just the additive constant will yield generators that are identical modulo an additive constant, and possibly a sign change. As a result,
changing the additive constant to generate multiple streams is a strategy to be avoided,
unless the output of the LCG is significantly mixed or combined with further sources of pseudorandomness.

\section{Using spectral data from MCGs}
\label{sec:MCGs}

The case of MCGs with power-of-two modulus is different from that of LCGs
because the maximum possible period is $m/4$~\citek[\S 3.2.1.2, Theorem C]{KnuACPII}. Thus, there are two
distinct orbits (remember that the state must be odd).
The nature of these orbits is, however, very different depending on whether the
multiplier is congruent to $5$ modulo $8$ or to $3$ modulo $8$: let us say such multipliers are
of \emph{type $5$} and \emph{type $3$}, respectively.

For multipliers of type $5$, each orbit is defined by the residue modulo $4$ of
the state (i.e., $1$ or $3$), whose value depends on the second-lowest
bit (this is a consequence of the fact that multipliers of type $5$
do not change the two lowest bits).  Thus, the
remaining upper bits (above the second) go through all possible $m/4$ values.  More importantly, the
lattice of points described by the upper bits is simply a translated version of
the lattice $\Lambda_d$ associated with the whole state, so the figures of merit we compute on
$\Lambda_d^*$ describe properties of the generator obtained by discarding the two lowest bits
from the state. Indeed, for every MCG of type $5$
there is an LCG with modulus $m/4$ that generates ``the same sequence'' if the
two low-order bits of every value produced by the MCG are ignored~\citek[\S 3.2.1.2, Exercise 9]{KnuACPII}.

For multipliers of type $3$, instead, each orbit is defined by the residue
modulo $8$ of the state: one orbit alternates between residues $1$ and $3$, and one
orbit alternates between $5$ and $7$ (because multipliers of type $3$
always leave the lowest bit and the third-lowest bit of the state unchanged). In this case, there is
no way to use the information we have about the lattice generated by
the whole state to obtain information about the lattice generated by the
part of state that is changing; indeed, there is again a correspondence with
an LCG, but the correspondence involves an alternating sign (again, see~\citek[\S 3.2.1.2, Exercise 9]{KnuACPII}).
For this reason, we (like L'Ecuyer~\cite{LEcTLCGDSGLS}) will consider only MCG multipliers of type $5$.

Note that $a$ and $-a \bmod m = m-a$ have different residue modulo $8$, but the same figures of merit~\citek[\S 3.2.1.2, Exercise 9]{KnuACPII}.
Moreover, in the MCG case
the lattice structure is invariant with respect to inversion modulo $m$, so for each
multiplier its inverse modulo $m$ has again the same figures of merit. In the end, for
each multiplier $a$ of maximum period $m/4$ there are three other related multipliers $a^{-1} \bmod m$,
$(-a) \bmod m$ and $ (- a^{-1}) \bmod m$ with the same figures of merit; of the four, two are of type $3$, and two of type $5$.
Therefore there is little lost in studying only multipliers of type 5.

\section{Search strategy}\label{sec:search}

For LCGs, only multipliers $a$ such that $a\bmod 8$ is either $1$ or $5$ achieve
full period~\citek[\S 3.2.1.2, Theorem A]{KnuACPII}, but we (like L'Ecuyer)
consider only the case of maximum potency, that is, the case when $a\bmod 8$ is $5$.
For MCGs, as we already discussed in Section~\ref{sec:MCGs}, we consider only multipliers of type $5$.
In the end, therefore, we consider in both cases (though for different reasons) only multipliers whose residue modulo $8$ is $5$. 

Following a suggestion in a paper by Entacher, Schell, and Uhl,~\cite{ESUELALGPS}, and
in the associated code,\footnote{\url{https://web.archive.org/web/20181128022136/http://random.mat.sbg.ac.at/results/karl/spectraltest/}} we
compute the figures of merit from $f_2$ to $f_8$ using the implementation of the
ubiquitous Lenstra--Lenstra--Lov{\'a}sz basis-reduction algorithm~\cite{LLL}
provided by Shoup's NTL library~\cite{ShoNTL}.
For $m=2^{64}$ and $m=^{128}$ we recorded in an output file all tested
multipliers whose minimum spectral score is at least $0.70$ (we used a lower
threshold for $m=2^{32}$).
Overall we sampled approximately $6.5\times10^{11}$ multipliers, enough to
ensure that for each pair of modulus and multiplier size reported, we recorded
at least one million multipliers.
(In several cases we recorded as many as $1.5$ million or even two million
multipliers.) As a sanity check, we also used the same software to test
multipliers of size $63$ for LCGs with $m=2^{128}$; as expected, in view of
Proposition~\ref{prop:bound} and its consequences, a random sample of well over
$10^{10}$ $63$-bit candidates revealed none whose minimum spectral score
is at least $0.70$.

In theory, the basis returned by the
algorithm is only necessarily made of shortest vectors, but using a precision parameter
$\delta=1-10^{-9}$ we found only very rarely a basis that was not made of such vectors:
we checked all multipliers we selected using the LatticeTester
tool,\footnote{\url{https://github.com/umontreal-simul/latticetester}} which
performs an exhaustive search after basis-reduction preprocessing, and almost all
approximated data we computed turned out to be exact; just a few cases (usually in high dimension)
were slightly off, which simply means that we spuriously stored a few candidates with minimum below $0.70$.
A similar strategy was also used by the LatMRG software.\cite{LECILSTMRLRNG}

Besides half-width and full-width multipliers, we searched for multipliers with up to two
bits more than half-width for $m=2^{32}$ and $m=2^{64}$, and up to seven bits more than half-width for $m=2^{128}$,
as well as multipliers of three-fourths width (24 bits for $m=2^{32}$, 48 bits for $m=2^{64}$, 96 bits for $m=2^{128}$),
because these are experimentally often as fast as smaller multipliers. Additionally, we provide $80$-bit
multipliers for $m=2^{128}$ because
such multipliers can be loaded by the ARM processor with just five instructions, and on an Intel processor one can use
a multiply instruction with an immediate $16$-bit value.

For small multipliers, we try to find candidates with a good $\lambda$: in particular,
we require that the second-most-significant bit be set. For
larger multipliers, we consider only spectral scores, as the effect of a good $\lambda$
becomes undetectable. Since when we consider $(w+c)$-bit multipliers we select candidates
larger than $2^{w+c-1}$, we have always $2^{c-1}\leq\lambda\leq 2^c$ for LCGs and  $2^{c}\leq\lambda\leq 2^{c+1}$ for MCGs.

\section{Selection}

We will now describe our selection criteria for the candidates computed during the search phase.
For each multiplier, we considered initially figures of merit up to dimension $8$, that
is, from $f_2$ to $f_8$, both for the
standard spectral test and for a lagged spectral test up to lag eight (for lack
of space, in our tables we show only the five unlagged scores $f_2$ through $f_6$). This procedure gives
us $56$ scores associated with each multiplier; criteria to summarize them
and propose a good multiplier are then a question of taste and personal preference.
Here, we decided to use the lagged scores to discard candidates with minimum
lagged score (over all lags and dimension) below the first quartile (i.e., we
discard three-fourths of the candidates). This approach eliminates candidates
with a mediocre lagged score, but leaves us with a large pool from which to
optimize the figures of merit from the standard spectral test.

Finally, even if our goal is to optimize figures of merit in low dimensions, to
avoid pathological behavior we furthermore computed the $24$ additional
(non-lagged) figures of merit $f_9$, $f_{10}$,~\ldots, $f_{32}$ and we discarded
a multiplier if a figure of merit up to dimension $32$ was below the threshold
$0.5$ (almost all are in the range $0.6-0.8$).

When examining the figures of merit of the spectral test up to
dimension $d$, typically multipliers are compared by their
\emph{minimum spectral score (up to dimension $d$)}, which is given by the \emph{minimum} figure of merit
over dimensions $2$ through $d$.  L'Ecuyer's paper~\cite{LEcTLCGDSGLS} uses the notation
$M_d(m,a)$ for this aggregate score for a generator with modulus $m$ and multiplier $a$.
We prefer to distinguish the minimum spectral scores of LCGs and MCGs, because the figures
of merit $f_d$ are computed differently for the two kinds of generator when the modulus is a power of two:
we use the notation
\[
\mathscr{M}^+_d(m,a) = \min_{2\leq i\leq d} f_i(m,a)
\]
to denote the minimum spectral score up to dimension $d$ for an LCG, and we use the notation
$\mathscr{M}^*_d(m,a)$ to denote the analogous score for an MCG.
The use of the minimum spectral score seems to have originated in the work of
Fishman and Moore~\cite{FiMEAMCRNG}, where, however, no motivation for this
choice is provided.

For small sets of multipliers, the minimum score does not pose particular problems. However,
once we have to select among a large database of candidates it sometimes sports pathological behavior.
Let us consider the following two multipliers for a $64$-bit LCG and their associated
scores from $f_2$ to $f_8$:
{\small
\begin{align*}
\mathtt{0xe2e19bb27190da6d}&	\quad 0.791216	 \quad 0.771300	 \quad 0.791569	 \quad 0.777944	 \quad 0.773526	 \quad 0.777463	 \quad 0.766073\\
\mathtt{0xe73d20db8e96d2cd}&	\quad 0.941271	 \quad 0.883251	 \quad 0.854317	 \quad 0.825078	 \quad 0.803654	 \quad 0.781546	 \quad 0.766043
\end{align*}
}
Because the $f_8$ score of the first multiplier, which is the minimum, is larger than the $f_8$ score of the
second multiplier, which is again the minimum, the minimum criterion would choose the first multiplier. This, clearly,
has no mathematical or empirical basis: the second multiplier has much better scores for $f_2$ through $f_7$, and
the difference between the $f_8$ scores, in the fifth decimal place, is entirely negligible.

To avoid such pathological behavior we use the following approach: first of all, we consider as equivalent
all multipliers whose minimum score is in the \emph{first millile} of our database, as the difference between the 
scores of such multipliers is practically negligible. Other criteria are possible, such as
a threshold on the difference with the best score in database, but a quantile criterion has the advantage
of being applicable uniformly on different data sets.

Within the first millile of minimum score, we then consider a secondary aggregate figure of merit:
\begin{definition}
Let $f_i(m,a)$, $2\leq i\leq d$, be the figures of merit of an LCG multiplier $a$ with modulus $m$. Then, the
\emph{harmonic spectral score (up to dimension $d$)} of $a$ with modulus $M$ is given by
\[
\mathscr H^+_d(m,a) = \frac{1}{H_{d-1}}\sum_{2\leq i\leq d} \frac{f_i(m,a)}{i-1},
\]
where $H_n=\sum_{k=1}^n\frac{1}{k}$ is the $n$-th harmonic number.\footnote{We have used script letters $\mathscr{M}$ and $\mathscr{H}$ to denote spectral scores so
that the harmonic spectral score function $\mathscr{H}_8$ will not be confused with the harmonic number $H_8$.}
Analogously, the notation $\mathscr H^*_d(m,a)$
denotes the harmonic spectral score (up to dimension $d$) for an MCG multiplier $a$ with modulus $m$.
\end{definition}
The effect of the harmonic spectral score is to weight each dimension progressively less,
using weights $1, 1/2, 1/3, \ldots, 1/(d-1)$, and the sum is normalized so that the score is always between $0$ and $1$.
In the example above, the harmonic score of the first multiplier is $0.782507$, whereas that of the second multiplier is $0.877164$.
Other within-millile criteria are possible (e.g., arithmetic mean, harmonic mean, etc.):  
however, Knuth argues that the importance of figures of merit decreases
with dimension, and that ``the values of $\nu_t$ for $t\geq 10$ seem to be of no
practical significance whatsoever''~\citek[\S 3.3.4]{KnuACPII}; similarly, L'Ecuyer
and Granger--Pich\'e, having to choose the congruential component for a combined generator~\cite{LEGCGCDF} (our main motivation),
suggest to consider only scores up to dimension eight. We thus prefer to privilege low dimensions.

To summarize: as a result of this multistep selection process, every multiplier we recommend has:
\begin{itemize}
\item a minimum lagged score in the top quartile of our original sample set (from the search procedure described in Section~\ref{sec:search}); 
\item no spectral score $f_9, f_{10}, f_{11}, \ldots, f_{32}$ below $0.5$; 
\item a minimum spectral score $\mathscr{M}^+_8(m,a)$ or $\mathscr{M}^*_8(m,a)$ that is in the first millile (top $0.1\%$) of remaining candidates; 
\item among all candidates in that first millile, the best harmonic score $\mathscr{H}^+_8(m,a)$ or $\mathscr{H}^*_8(m,a)$.
\end{itemize}

We remark that on smaller sets of multipliers (e.g., $16$-bits multipliers for $32$-bit generators) the first millile contains just one element; in such cases
the harmonic score is not involved in the selection.

Tables~\ref{tab:lcg32}--\ref{tab:mcg128}
contain the best (by the criterion above) multipliers we found, and other similar multipliers are available online.
All multipliers we provide are \emph{Pareto optimal} for our dataset: that is,
for each type, modulus, and size there is no other multiplier we examined that is
at least as good on both scores, and strictly improves one.

\section{Conclusions}

We have presented new tables of multipliers for LCGs and MCGs from size $32$ bits up to size $128$ bits.
Based on the observation that smaller multipliers can lead to faster generators, we provide multipliers
of different bit sizes. Beside using the standard minimum criterion up to dimension eight, to avoid pathological behavior in the
presence of several multipliers with practically indistinguishable minima we introduced a new disambiguation
score, the harmonic score, which privileges low dimensions. While we optimize for lower dimensions, we
use thresholding to avoid multipliers that are pathological in high dimension or in lagged sequences.
The complete database of multipliers from our search procedure (prior to the specific selection procedure
that we applied to produce the tables) is available online.

We remark that good spectral scores are known to be correlated with
success in certain statistical tests, and in particular with the collision
test~\citek[\S 3.3.2.I]{KnuACPII} and the birthday-spacing test~\citek[\S
3.3.2.J]{KnuACPII}. We have verified empirically that in several cases our
multipliers provide LCGs that in isolation perform better than previous
proposals (e.g., better than Knuth's MMIX LCG~\cite{KnuACPII} with multiplier
$6364136223846793005$). However, there is currently no
mathematical theory that can prove whether these improvements percolate to the
output once other PRNGs are mixed in, or when using output functions with good mixing properties, as in the
LXM case. Improving measurably the quality of a component is a best practice in software
design, in particular if the improvement comes at no cost, as in this case. We leave to
future work the development of theoretical or empirical tests that can detect
the improvement in spectral scores in the LCG component of a combined generator.

\newbox\tempa
\newbox\tempb
\newcommand\headertowidth[3]{{\setbox\tempa\hbox{\tt #2}\setbox\tempb\hbox{\tt #3}\hbox to \wd\tempb{\hfil\hbox to \wd\tempa{\hfil#1\hfil}}}}

\begin{sidewaystable}
\centering
\renewcommand{\arraystretch}{0.9}
\setlength{\tabcolsep}{6pt}
\begin{tabular}{@{}rrrrrrrrrl@{}}
\multicolumn{1}{@{}c}{Bits}&\multicolumn{1}{c}{$\mathscr M^+_8(m,a)$}&\multicolumn{1}{c}{$\mathscr H^+_8(m,a)$}&\multicolumn{1}{c}{\headertowidth{$a$}{0x00000000}{0x00000000000000000000000000000000}}&\multicolumn{1}{c}{$f_2$}&\multicolumn{1}{c}{$f_3$}&\multicolumn{1}{c}{$f_4$}&\multicolumn{1}{c}{$f_5$}&\multicolumn{1}{c}{$f_6$}&\multicolumn{1}{c@{}}{\headertowidth{\llap{$\lambda$ }}{$0.0\times10^{00}$}{$0.0\times10^{00}$}}\\\midrule
$16$
	&	$0.6690$	&	$0.7444$	&	\texttt{0xd9f5}	&	$0.7923$	&	$0.7541$	&	$0.6869$	&	$0.6690$	&	$0.6919$	&	$\phantom{00}0.85$\\
$17$
	&	$0.6403$	&	$0.6588$	&	\texttt{0x1dab5}	&	$0.6652$	&	$0.6468$	&	$0.6555$	&	$0.6455$	&	$0.6403$	&	$\phantom{00}1.85$\\
$18$
	&	$0.6988$	&	$0.7362$	&	\texttt{0x3d575}	&	$0.7304$	&	$0.7186$	&	$0.7855$	&	$0.7422$	&	$0.7222$	&	$\phantom{00}3.83$\\
$24$
	&	$0.7224$	&	$0.8371$	&	\texttt{0xc083c5}	&	$0.9397$	&	$0.8395$	&	$0.7224$	&	$0.7268$	&	$0.7976$	&	$192.51$\\
$32$
	&	$0.7591$	&	$0.8038$	&	\texttt{0x915f77f5}	&	$0.7900$	&	$0.8641$	&	$0.7932$	&	$0.7713$	&	$0.8325$	&	$3.7\times10^{4}$\\

\end{tabular}
\caption{\label{tab:lcg32}Good multipliers for LCGs with $m=2^{32}$.}
\vspace*{.2cm}
\begin{tabular}{@{}rrrrrrrrrl@{}}
\multicolumn{1}{@{}c}{Bits}&\multicolumn{1}{c}{$\mathscr M^*_8(m,a)$}&\multicolumn{1}{c}{$\mathscr H^*_8(m,a)$}&\multicolumn{1}{c}{\headertowidth{$a$}{0x00000000}{0x00000000000000000000000000000000}}&\multicolumn{1}{c}{$f_2$}&\multicolumn{1}{c}{$f_3$}&\multicolumn{1}{c}{$f_4$}&\multicolumn{1}{c}{$f_5$}&\multicolumn{1}{c}{$f_6$}&\multicolumn{1}{c@{}}{\headertowidth{\llap{$\lambda$ }}{$0.0\times10^{00}$}{$0.0\times10^{00}$}}\\\midrule
$15$
	&	$0.6814$	&	$0.7502$	&	\texttt{0x72ed}	&	$0.8356$	&	$0.6814$	&	$0.6978$	&	$0.7213$	&	$0.6909$	&	$\phantom{00}0.90$\\
$16$
	&	$0.6610$	&	$0.7462$	&	\texttt{0xecc5}	&	$0.7066$	&	$0.9008$	&	$0.6610$	&	$0.7483$	&	$0.7002$	&	$\phantom{00}1.85$\\
$17$
	&	$0.6623$	&	$0.7497$	&	\texttt{0x1e92d}	&	$0.8419$	&	$0.6851$	&	$0.6770$	&	$0.7050$	&	$0.6623$	&	$\phantom{00}3.82$\\
$18$
	&	$0.6814$	&	$0.8266$	&	\texttt{0x39e2d}	&	$0.9021$	&	$0.8643$	&	$0.7004$	&	$0.8121$	&	$0.7936$	&	$\phantom{00}7.24$\\
$24$
	&	$0.7064$	&	$0.8344$	&	\texttt{0xe47135}	&	$0.9293$	&	$0.8651$	&	$0.7064$	&	$0.7519$	&	$0.7635$	&	$456.88$\\
$32$
	&	$0.7470$	&	$0.8649$	&	\texttt{0x93d765dd}	&	$0.9633$	&	$0.8674$	&	$0.7774$	&	$0.7826$	&	$0.7929$	&	$7.6\times10^{4}$\\

\end{tabular}
\caption{\label{tab:mcg32}Good multipliers for MCGs with $m=2^{32}$.}
\vspace*{.2cm}
\begin{tabular}{@{}rrrrrrrrrl@{}}
\multicolumn{1}{@{}c}{Bits}&\multicolumn{1}{c}{$\mathscr M^+_8(m,a)$}&\multicolumn{1}{c}{$\mathscr H^+_8(m,a)$}&\multicolumn{1}{c}{\headertowidth{$a$}{0x0000000000000000}{0x00000000000000000000000000000000}}&\multicolumn{1}{c}{$f_2$}&\multicolumn{1}{c}{$f_3$}&\multicolumn{1}{c}{$f_4$}&\multicolumn{1}{c}{$f_5$}&\multicolumn{1}{c}{$f_6$}&\multicolumn{1}{c@{}}{\headertowidth{\llap{$\lambda$ }}{$0.0\times10^{00}$}{$0.0\times10^{00}$}}\\\midrule
$32$
	&	$0.7571$	&	$0.8364$	&	\texttt{0xf9b25d65}	&	$0.9077$	&	$0.8252$	&	$0.7882$	&	$0.7762$	&	$0.7837$	&	$\phantom{00}0.98$\\
$33$
	&	$0.7525$	&	$0.8105$	&	\texttt{0x18a44074d}	&	$0.8694$	&	$0.7932$	&	$0.7786$	&	$0.7525$	&	$0.7609$	&	$\phantom{00}1.54$\\
$34$
	&	$0.7573$	&	$0.8405$	&	\texttt{0x3af78c385}	&	$0.9299$	&	$0.8115$	&	$0.7843$	&	$0.7700$	&	$0.7573$	&	$\phantom{00}3.69$\\
$48$
	&	$0.7590$	&	$0.8821$	&	\texttt{0xc2ec33ef97a5}	&	$0.9947$	&	$0.8554$	&	$0.8299$	&	$0.7617$	&	$0.7887$	&	$5.0\times10^{4}$\\
$64$
	&	$0.7602$	&	$0.8992$	&	\texttt{0xd1342543de82ef95}	&	$0.9586$	&	$0.9375$	&	$0.8708$	&	$0.8223$	&	$0.8204$	&	$3.5\times10^{9}$\\

\end{tabular}
\caption{\label{tab:lcg64}Good multipliers for LCGs with $m=2^{64}$.}
\vspace*{.2cm}
\begin{tabular}{@{}rrrrrrrrrl@{}}
\multicolumn{1}{@{}c}{Bits}&\multicolumn{1}{c}{$\mathscr M^*_8(m,a)$}&\multicolumn{1}{c}{$\mathscr H^*_8(m,a)$}&\multicolumn{1}{c}{\headertowidth{$a$}{0x0000000000000000}{0x00000000000000000000000000000000}}&\multicolumn{1}{c}{$f_2$}&\multicolumn{1}{c}{$f_3$}&\multicolumn{1}{c}{$f_4$}&\multicolumn{1}{c}{$f_5$}&\multicolumn{1}{c}{$f_6$}&\multicolumn{1}{c@{}}{\headertowidth{\llap{$\lambda$ }}{$0.0\times10^{00}$}{$0.0\times10^{00}$}}\\\midrule
$32$
	&	$0.7490$	&	$0.7905$	&	\texttt{0xe817fb2d}	&	$0.8268$	&	$0.7765$	&	$0.7803$	&	$0.7535$	&	$0.7490$	&	$\phantom{00}1.81$\\
$33$
	&	$0.7630$	&	$0.8401$	&	\texttt{0x1e85bbd25}	&	$0.8499$	&	$0.9149$	&	$0.8147$	&	$0.7971$	&	$0.7899$	&	$\phantom{00}3.82$\\
$34$
	&	$0.7524$	&	$0.8670$	&	\texttt{0x34edd34ad}	&	$0.9749$	&	$0.8506$	&	$0.7789$	&	$0.8145$	&	$0.7550$	&	$\phantom{00}6.62$\\
$48$
	&	$0.7722$	&	$0.8894$	&	\texttt{0xbdcdbb079f8d}	&	$0.9855$	&	$0.8937$	&	$0.7973$	&	$0.8466$	&	$0.7867$	&	$9.7\times10^{4}$\\
$64$
	&	$0.7584$	&	$0.8797$	&	\texttt{0xf1357aea2e62a9c5}	&	$0.9705$	&	$0.8444$	&	$0.8415$	&	$0.7928$	&	$0.8202$	&	$8.1\times10^{9}$\\

\end{tabular}
\caption{\label{tab:mcg64}Good multipliers for MCGs with $m=2^{64}$.}
\end{sidewaystable}

\begin{sidewaystable}[p]
\centering
\renewcommand{\arraystretch}{0.9}
\setlength{\tabcolsep}{6pt}
\begin{tabular}{@{}rrrrrrrrrl@{}}
\multicolumn{1}{@{}c}{Bits}&\multicolumn{1}{c}{$\mathscr M^+_8(m,a)$}&\multicolumn{1}{c}{$\mathscr H^+_8(m,a)$}&\multicolumn{1}{c}{\headertowidth{$a$}{0x00000000000000000000000000000000}{0x00000000000000000000000000000000}}&\multicolumn{1}{c}{$f_2$}&\multicolumn{1}{c}{$f_3$}&\multicolumn{1}{c}{$f_4$}&\multicolumn{1}{c}{$f_5$}&\multicolumn{1}{c}{$f_6$}&\multicolumn{1}{c@{}}{\headertowidth{\llap{$\lambda$ }}{$0.0\times10^{00}$}{$0.0\times10^{00}$}}\\\midrule
$64$
	&	$0.7596$	&	$0.8630$	&	\texttt{0xfc0072fa0b15f4fd}	&	$0.9161$	&	$0.9285$	&	$0.7850$	&	$0.8158$	&	$0.7596$	&	$\phantom{00}0.98$\\
$65$
	&	$0.7752$	&	$0.8882$	&	\texttt{0x1ed5301a365eced85}	&	$0.9682$	&	$0.9138$	&	$0.8134$	&	$0.8192$	&	$0.8036$	&	$\phantom{00}1.93$\\
$66$
	&	$0.7579$	&	$0.8810$	&	\texttt{0x3cf736df8904b7285}	&	$0.9869$	&	$0.8444$	&	$0.8468$	&	$0.8096$	&	$0.7698$	&	$\phantom{00}3.81$\\
$67$
	&	$0.7592$	&	$0.8982$	&	\texttt{0x77808d182e9136c35}	&	$0.9721$	&	$0.9324$	&	$0.8682$	&	$0.8175$	&	$0.7995$	&	$\phantom{00}7.47$\\
$68$
	&	$0.7774$	&	$0.8656$	&	\texttt{0xc0c5c8bdde1eae3c5}	&	$0.9675$	&	$0.7774$	&	$0.8283$	&	$0.8138$	&	$0.8222$	&	$\phantom{0}12.05$\\
$69$
	&	$0.7739$	&	$0.8665$	&	\texttt{0x1ffc98ea48aaa4c915}	&	$0.9852$	&	$0.7763$	&	$0.8094$	&	$0.8298$	&	$0.7739$	&	$\phantom{0}31.99$\\
$70$
	&	$0.7551$	&	$0.8845$	&	\texttt{0x3d0e4c1575cfe11085}	&	$0.9768$	&	$0.9173$	&	$0.7864$	&	$0.7937$	&	$0.7959$	&	$\phantom{0}61.06$\\
$71$
	&	$0.7603$	&	$0.8844$	&	\texttt{0x774236e651b3d34775}	&	$0.9696$	&	$0.9168$	&	$0.8108$	&	$0.7603$	&	$0.7662$	&	$119.26$\\
$72$
	&	$0.7592$	&	$0.8830$	&	\texttt{0xce12be0ad9384349d5}	&	$0.9548$	&	$0.9322$	&	$0.8139$	&	$0.7592$	&	$0.8244$	&	$206.07$\\
$80$
	&	$0.7651$	&	$0.8642$	&	\texttt{0xe14777aac3fc617a34ad}	&	$0.9779$	&	$0.8050$	&	$0.7993$	&	$0.7744$	&	$0.8040$	&	$5.8\times10^{4}$\\
$96$
	&	$0.7771$	&	$0.8903$	&	\texttt{0xdfe1956283473c8e63b49445}	&	$0.9599$	&	$0.9367$	&	$0.8193$	&	$0.8042$	&	$0.8205$	&	$3.8\times10^{9}$\\
$128$
	&	$0.7650$	&	$0.8878$	&	\texttt{0xdb36357734e34abb0050d0761fcdfc15}	&	$0.9849$	&	$0.8546$	&	$0.8084$	&	$0.8527$	&	$0.8209$	&	$1.6\times10^{19}$\\

\end{tabular}
\caption{\label{tab:lcg128}Good multipliers for LCGs with $m=2^{128}$.}
\vspace*{1cm}
\begin{tabular}{@{}rrrrrrrrrl@{}}
\multicolumn{1}{@{}c}{Bits}&\multicolumn{1}{c}{$\mathscr M^*_8(m,a)$}&\multicolumn{1}{c}{$\mathscr H^*_8(m,a)$}&\multicolumn{1}{c}{\headertowidth{$a$}{0x00000000000000000000000000000000}{0x00000000000000000000000000000000}}&\multicolumn{1}{c}{$f_2$}&\multicolumn{1}{c}{$f_3$}&\multicolumn{1}{c}{$f_4$}&\multicolumn{1}{c}{$f_5$}&\multicolumn{1}{c}{$f_6$}&\multicolumn{1}{c@{}}{\headertowidth{\llap{$\lambda$ }}{$0.0\times10^{00}$}{$0.0\times10^{00}$}}\\\midrule
$63$
	&	$0.7547$	&	$0.8649$	&	\texttt{0x7e91d554f7f50a65}	&	$0.9202$	&	$0.9217$	&	$0.7759$	&	$0.8492$	&	$0.7783$	&	$\phantom{00}0.99$\\
$64$
	&	$0.7550$	&	$0.8819$	&	\texttt{0xdefba91144f2b375}	&	$0.9502$	&	$0.9370$	&	$0.7973$	&	$0.8203$	&	$0.7997$	&	$\phantom{00}1.74$\\
$65$
	&	$0.7558$	&	$0.8722$	&	\texttt{0x1f3e451e8032e03cd}	&	$0.9951$	&	$0.8408$	&	$0.7558$	&	$0.7725$	&	$0.8285$	&	$\phantom{00}3.91$\\
$66$
	&	$0.7734$	&	$0.8722$	&	\texttt{0x32afe9369a79aaf95}	&	$0.9180$	&	$0.8897$	&	$0.8898$	&	$0.8057$	&	$0.8011$	&	$\phantom{00}6.34$\\
$67$
	&	$0.7578$	&	$0.8648$	&	\texttt{0x79ee767d1056fa525}	&	$0.9613$	&	$0.8551$	&	$0.8076$	&	$0.7891$	&	$0.7578$	&	$\phantom{0}15.24$\\
$68$
	&	$0.7622$	&	$0.8752$	&	\texttt{0xee8f25664dcc71505}	&	$0.9447$	&	$0.8999$	&	$0.8044$	&	$0.8417$	&	$0.7984$	&	$\phantom{0}29.82$\\
$69$
	&	$0.7570$	&	$0.8728$	&	\texttt{0x1c9539b04653db147d}	&	$0.9577$	&	$0.8853$	&	$0.8133$	&	$0.8074$	&	$0.7758$	&	$\phantom{0}57.17$\\
$70$
	&	$0.7575$	&	$0.8725$	&	\texttt{0x3a6de753da52a34cdd}	&	$0.9645$	&	$0.9030$	&	$0.7793$	&	$0.7827$	&	$0.7575$	&	$116.86$\\
$71$
	&	$0.7658$	&	$0.8753$	&	\texttt{0x654421647a648c4eb5}	&	$0.9324$	&	$0.9166$	&	$0.8357$	&	$0.8236$	&	$0.7658$	&	$202.53$\\
$72$
	&	$0.7627$	&	$0.8693$	&	\texttt{0xeee2a2bb3ec40a671d}	&	$0.9875$	&	$0.8125$	&	$0.7791$	&	$0.8248$	&	$0.7627$	&	$477.77$\\
$80$
	&	$0.7610$	&	$0.8727$	&	\texttt{0xd49d0e59c6b198994025}	&	$0.9591$	&	$0.8845$	&	$0.8281$	&	$0.7708$	&	$0.7860$	&	$1.1\times10^{5}$\\
$96$
	&	$0.7566$	&	$0.8721$	&	\texttt{0xb540ecedd0778e2d651421a5}	&	$0.9654$	&	$0.8510$	&	$0.8244$	&	$0.8128$	&	$0.7875$	&	$6.1\times10^{9}$\\
$128$
	&	$0.7571$	&	$0.8799$	&	\texttt{0xaadec8c3186345282b4e141f3a1232d5}	&	$0.9869$	&	$0.8706$	&	$0.8056$	&	$0.7838$	&	$0.7571$	&	$2.5\times10^{19}$\\

\end{tabular}
\caption{\label{tab:mcg128}Good multipliers for MCGs with $m=2^{128}$.}
\end{sidewaystable}

\bibliography{biblio,mult}

\end{document}